\documentclass[aps,pre,twocolumn,superscriptaddress,showpacs,preprintnumbers]{revtex4-2} 

\usepackage{hyperref}
\usepackage{natbib}
\usepackage{etoolbox}
\apptocmd{\thebibliography}{\raggedright}{}{} 

\usepackage[utf8]{inputenc}
\usepackage{subfiles}
\usepackage{amsfonts}
\usepackage{graphicx}
\usepackage{physics}
\usepackage{float}
\usepackage{amsmath}
\usepackage{mathtools}
\usepackage[english]{babel}
\usepackage{csquotes}

\usepackage[draft,inline,nomargin]{fixme} \fxsetup{theme=color}
\FXRegisterAuthor{cp}{acp}{\color{teal}CP}
\FXRegisterAuthor{gs}{ags}{\color{teal}GS}

\usepackage[linecolor=red!30]{todonotes}
\tikzstyle{notestyleraw} = [
	draw=black!0,
	line width=0.8pt,
	fill=red!10,
	text=black,
    text width=1.5cm,
	inner sep=0.8 ex
]

\DeclareMathOperator{\C}{C}
\DeclareMathOperator{\Erg}{Erg}
\DeclareMathOperator{\PDF}{PDF}
\DeclareMathOperator{\CDF}{CDF}
\DeclareMathOperator{\eCDF}{eCDF}

\newcommand{\KS}{\Delta_{\text{KS}} }

\newcommand{\psiz}{\ket*{\psi_0}}
\newcommand{\psit}{\ket*{\psi_t}}

\newcommand{\vphi}{\varphi}

\newcommand{\ketn}[1]{\ket*{K_{#1} }}

\newcommand{\ant}[2]{\alpha_{#1\,#2}}

\newcommand{\uba}{Universidad de Buenos Aires, Facultad de Ciencias Exactas y Naturales, Departamento de Física. Buenos Aires 1428, Argentina}
\newcommand{\ifiba}{CONICET - Universidad de Buenos Aires, Instituto de Física de Buenos Aires (IFIBA). Buenos Aires 1428, Argentina}
\newcommand{\unam}{Instituto de Física, Universidad Nacional Autónoma de México, Ciudad de México 01000, Mexico
}
\begin{document}
\title{Exploring quantum ergodicity of unitary evolution through the Krylov approach}

    \author{Gast\'on F. Scialchi}
    \email[E--mail address: ]{gscialchi@df.uba.ar}
    \affiliation{\uba}
    \affiliation{\ifiba}

    \author{Augusto J. Roncaglia}
    \affiliation{\uba}
    \affiliation{\ifiba}

    \author{Carlos Pineda}
    \affiliation{\unam}

    \author{Diego A. Wisniacki}
    \email[E--mail address: ]{wisniacki@df.uba.ar}
    \affiliation{\uba}
    \affiliation{\ifiba}

    \begin{abstract}
        In recent years there has been growing interest in characterizing the complexity of quantum evolutions of interacting many-body systems.
        When a time-independent Hamiltonian governs the dynamics, Krylov complexity has emerged as a powerful tool.
        For unitary evolutions like kicked systems or Trotterized dynamics, a similar formulation based on the Arnoldi approach has been proposed yielding a new notion of quantum ergodicity  [P. Suchsland, R. Moessner, and P. W. Claeys, Phys. Rev. B 111, 014309 (2025)].
        In this work, we show that this formulation is robust for observing the transition from integrability to chaos in both autonomous and kicked systems. Examples from random matrix theory and spin chains are shown in this paper.
    \end{abstract}

\maketitle
\section{Introduction} 
\label{sec:introduction}

Characterizing the complexity of quantum evolutions is a relevant topic nowadays, especially with the advances in experimental control of many-body systems \cite{schweigler2017experimental, zhang2017observation}. Balancing control with complexity is often challenging, underscoring its significance in emerging quantum technologies. Therefore, having robust and computable measures of quantum chaoticity is crucial for grasping the evolution of many-body systems.

In the context of systems whose evolution is described by a time-independent Hamiltonian, the Krylov complexity has emerged as a promising approach for addressing the inherent complexity of their evolution. This approach, which can be applied for both states and operators, has gained considerable attention in recent years \cite{parker2019, Rabinovici2021, Rabinovici_2022_integrability, Balasubramanian_2022, PhysRevE.107.024217, PhysRevE.109.054209, domingo2023quantum}.
In this scenario, the evolution is mapped onto the dynamics of a single-particle wavefunction in the so-called Krylov subspace,
which is governed by a tight-binding model.
In terms of this mapping, a complexity measure is defined as the average position of the particle within the Krylov basis.
The concept of ``spread complexity" was proposed as a means of quantifying the extent of the wavefunction spreading over an arbitrary reference basis. It was also argued that the Krylov basis minimizes this spreading, making the Krylov complexity a particularly suitable measure for characterizing chaos~\cite{Balasubramanian_2022, PhysRevE.109.054209}.

For non autonomous systems, such as kicked, Trotterized, or time-dependent Hamiltonian systems,
notions of complexity based on the Krylov construction
have been less extensively studied~\cite{nizami2023krylov, suchsland2023krylov,nizami2024spreadcomplexityquantumchaos}.
In this context, it has been shown that the superoperator governing the Heisenberg evolution of operators assumes an upper Hessenberg matrix form in the Krylov basis.
In Ref. \cite{suchsland2023krylov}, it is conjectured that for unitary circuit dynamics, the aforementioned superoperator will tend towards an ergodic form where it becomes purely lower diagonal in the Krylov basis. This ergodic hypothesis has been demonstrated in a Floquet circuit. However, its validation in other systems is of great significance, as it can be employed in the characterization of generic unitary evolutions, which are essential for simulations of quantum systems in the NISQ era \cite{daley2022practical,berthusen2022quantum}.

The aim of this work is to investigate whether this complexity measure based on the ergodic form is related with the usual measures of quantum chaos, i.e., to verify whether the condition of maximum ergodicity is connected to the statistical measures of the eigenvalues and eigenfunctions as predicted by random matrix theory (RMT).
The first step was to adapt the formalism from Ref.~\cite{suchsland2023krylov} on operators to the case of state evolution,
and define an ergodicity measure based on the previously mentioned ergodic form.
We studied how this new measure of ergodicity behaves for an autonomous evolution given by a Hamiltonian, chosen from a random matrix ensemble that parametrically transitions from Poisson statistics to Wigner-Dyson statistics.
Then we studied the same phenomenon in an Ising-type spin chain with transverse
and longitudinal magnetic fields, for which we have considered the autonomous
and Trotterized evolutions. We found that, in all these models, the ergodicity
measure based on the Arnoldi iteration is very useful for determining the
transition between integrability and chaos. This seemingly universal behavior
establishes this form of ergodicity as a fundamental quantity for future
analyses of many-body systems.

The paper is organized as follows. In Sec. \ref{sec:arnoldi}, we present the Arnoldi approach to complexity, describing how the Arnoldi form is obtained for unitary evolutions, and defining a distance measure relative to the maximum ergodicity.
In Sec. \ref{sec:chaos_measures}, we give a brief summary of the usual measures of quantum chaos used in this work
against which the ergodicity measure will be compared.
In Sec. \ref{sec:results}, we describe the models used in this work—a parametric random matrix model and a spin chain with transverse and longitudinal magnetic fields—
and we analyze the behavior of the complexity measure based on the Arnoldi form.
We present our conclusions in Sec. \ref{sec:conclusions}.

\section{Krylov approach to unitary evolution} \label{sec:arnoldi} 
The Arnoldi method is an iterative algorithm that is commonly used to
approximate the eigenvalues and eigenvectors of large matrices
\cite{arnoldi1951principle,bai2005soar,bellalij2007convergence}.  In this
context, it also allows to generate the Krylov subspace for unitary dynamics,
although it has also been used for the dynamics of open quantum systems
\cite{Bhattacharya:2022gbz}.  In particular, in Ref.~\cite{suchsland2023krylov}
the Arnoldi method has been applied to study the evolution due to unitary
quantum circuits in Krylov space. Bellow, we will briefly describe some
properties of the unitary operator that generates the evolution in the Krylov
basis, and we define a measure of ergodicity based on these results.
\subsection{Arnoldi iteration method}
\label{sec:arnoldi_method}

In this section we show how to represent the unitary operator $U$, which drives the evolution of a quantum system, in the Krylov basis. The procedure that we follow is an extension of the method derived in ~\cite{suchsland2023krylov} that was applied to the study of unitary circuits.

Given an initial state $\psiz$ and a unitary $U$, the full dynamics is characterized by the sequence of states $\psit = U^t \psiz$ with $t\in \mathbb N_0$.
The  Krylov basis $\{\ketn{0},\ketn{1}, \ldots\}$, with $\braket{K_n}{K_m}=\delta_{nm}$, is defined by a Gram-Schmidt orthonormalization of the sequence $\{\ket{\psi_0},\ket{\psi_1}, \ldots\}$.
Here $\ket{K_0}=\ket{\psi_0}$, and the remaining elements of the basis can be inductively defined as
\begin{equation}
    b_n \ketn{n} = U \ketn{n-1} - \sum_{l=0}^{n-1} \mel{K_l}{U}{K_{n-1}} \ketn{l}
    \label{eq:arnoldi_it}.
\end{equation}
Written in the Krylov basis, the unitary $U$ attains an upper Hessenberg form, since $\mel{K_m}{U}{K_n} = 0$ for any $m > n + 1$.
Moreover, the operator $U$ can be completely parametrized by the sequences
\begin{eqnarray}
    b_n &=& \mel{K_n}{U}{K_{n-1}}, \quad
    a_n = \mel{K_n}{U}{K_n},\nonumber \\
    c_n &=& \mel{K_0}{U}{K_n}.
    \label{eq:sequences}
\end{eqnarray}
This can be derived by expanding the elements of the Krylov basis in terms of the evolved states $\psit = U^t \psiz$
\begin{equation}
    \ketn{n} = \sum_{t=0}^n \alpha_{n,t} \psit
    \label{eq:kn_psit}
\end{equation}
and conversely
\begin{equation}
    \psit = \sum_{n=0}^t \beta_{t,n} \ketn{n}
    \label{eq:psit_kn}.
\end{equation}
Since $\braket{\psi_t}{K_n} = 0$ $\forall t < n$
one obtains
$\mel{K_m}{U}{K_n} = \ant{m}{0}^* c_n$ $\forall\, m\leq n$.
In particular $a_m = \ant{m}{0}^* c_m$,
which yields
\begin{equation}
    \mel{K_m}{U}{K_n} = \frac{a_m}{c_m} c_n \quad \forall\, m\leq n
    \label{eq:kmukn_ancn},
\end{equation}
meaning that all matrix elements are written in terms of the sequences \eqref{eq:sequences}.
Note that while the sequence $b_n$ is real and positive, the sequences $a_n$ and $c_n$ are generally complex.
Equation~\eqref{eq:kn_psit} implies that the Krylov states $\ketn{n}$ can also be represented in terms of the application of some polynomials $p_n(U)$,
\begin{equation}
    p_n(U) = \sum_{t=0}^n \alpha_{n,t} U^t
    \label{eq:pn},
\end{equation}
to the initial state: $\ketn{n} = p_n(U) \psiz$.
As the Krylov basis is orthonormal $\braket{K_m}{K_n} = \delta_{m n}$,
so are the polynomials with respect to some weight function $w$
\begin{equation}
    \frac{1}{2\pi} \int d\vphi\, w(\vphi) p_m(z)^* p_n(z) = \delta_{m n},
    \label{eq:pmpn_ort_measure}
\end{equation}
    where $w(\vphi)$ is defined through
\begin{equation}
    \frac{1}{2\pi} \int d\vphi\, w(\vphi) f(z) = \ev{f(U)}{\psi_0},
\end{equation}
with $z=e^{i\vphi}$ the eigenvalues of $U$, i.e. $U \ket{\vphi} = e^{i\vphi} \ket{\vphi}$,
and the integrations are done over $\psiz$'s support on the unitary's spectrum.
In term of these polynomials, the sequences \eqref{eq:sequences} can be expressed as
\begin{eqnarray}
    a_n &=& \frac{1}{2\pi} \int d\vphi\, w(\vphi) e^{i\vphi} \abs{p_n(z)}^2, \nonumber\\
    b_n &=& \frac{1}{2\pi} \int d\vphi\, w(\vphi) e^{i\vphi} p_n(z)^* p_{n-1}(z), \nonumber\\
    c_n &=& \frac{1}{2\pi} \int d\vphi\, w(\vphi) e^{i\vphi} p_n(z)
    \label{eq:sequences_pn}.
\end{eqnarray}
Notice that all the matrix elements are bounded, since $|{U}_{m,n}|\leq 1$,
so the Arnoldi sequences contrast with the Lanczos coefficients that arise from continuous-time evolution
in that the latter are generally not bounded.
In fact, a linear growth of the sequence $b_n$ for large $n$ was originally proposed as a signature of chaotic behavior \cite{parker2019}.

\subsection{A measure of ergodicity}
\label{sec:ergodicity}
Operator growth is a ubiquitous probe of quantum chaotic
dynamics~\cite{Garcia-Mata:2023,PRXQuantum.5.010201,swingle2018unscrambling,nandy2024quantum}.
In fact, a ``universal operator growth hypothesis'' has been proposed for
autonomous Hamiltonian dynamics, stating that maximally chaotic systems posses
a particular Krylov space structure, leading to an exponential growth of the
operator complexity measure known as Krylov complexity~\cite{parker2019}.
This notion has been extended to unitary circuit dynamics,
where ``maximally ergodic Krylov dynamics'' was found in the Krylov space of dual-unitary circuits.
In these systems, the Krylov dynamics is such that
$\mathcal{U} O_n = O_{n+1}$ after a single circuit step,
meaning that the Krylov space is explored most efficiently.
Here $O_n$ refers to an operator Krylov basis element,
and $\mathcal{U}$ to the unitary superoperator such that $O(t) = U^{\dagger t} O U^t$
for any observable $O$.
As a consequence, the operator Krylov complexity grows linearly in time,
which is the maximal growth allowed by unitarity.
Such a regime was proposed not to be particular to dual-unitary dynamics,
but to be a universal feature of chaotic unitary circuits \cite{suchsland2023krylov}.

In this section, we will show the emergence of the maximally ergodic regime for
the Schrödinger evolution of states, and we define a probe for ``ergodicity''
derived from the particular structure the Krylov space attains in that regime.
Furthermore, we will argue that this behavior is not only a characteristic of
unitary circuit dynamics, but is a phenomenon also present in autonomous
unitary evolutions.

The maximally ergodic regime identified in Ref.~\cite{suchsland2023krylov} arises from the asymptotic behavior for the polynomials \eqref{eq:pn} which,
under some regularity conditions upon the weight function $w$,
satisfy the asymptotic behavior \cite{szego}
\begin{equation}
    p_n(z) \approx \frac{z^n}{\sqrt{w(\vphi)}}
    \quad\text{for } n \gg 1
    \label{eq:pn_lim}.
\end{equation}
As a consequence, for large $n$ the matrix elements of $U$ in the Krylov basis
fulfill the following three properties~\cite{suchsland2023krylov}:

(i) The sequences \eqref{eq:sequences_pn} in  this limit are:
\begin{eqnarray}
    a_n &&= \frac{1}{2\pi} \int d\vphi\, e^{i\vphi} = 0,\quad
    b_n = \frac{1}{2\pi} \int d\vphi\, = 1, \nonumber \\
    c_n &&= \frac{1}{2\pi} \int d\vphi\, \sqrt{w(\vphi)} e^{i\vphi(n+1)} = f_{n+1}^* \to
    0
    \label{eq:sequences_lim},
\end{eqnarray}
where $f_{n}$ is the $n$th Fourier coefficient of $\sqrt{w(\vphi)}$.
Thus, $U$ becomes purely lower diagonal in the Krylov basis,
implying $U \ketn{n} = \ketn{n+1}$.

(ii) The autocorrelators for the Krylov states $\C^{(n)}_t = \ev{U^t}{K_n}$,
which can be written as
\begin{equation}
    \C^{(n)}_t = \frac{1}{2\pi} \int d\vphi\, w(\vphi) e^{i t\vphi} \abs{p_n(z)}^2
    \label{eq:cn_t},
\end{equation}
attain the form
\begin{equation}
    \C^{(n)}_t = \frac{1}{2\pi} \int d\vphi\, e^{i t\vphi} = \delta_{t 0}
    \label{eq:cn_t_lim}
\end{equation}
 such that their autocorrelations decay in a single time step.

(iii) The discrete-time Fourier transforms of these autocorrelations $\C^{(n)}_\omega$ become constant,
which follows straightforwardly from \eqref{eq:cn_t_lim}.

A relevant aspect to keep in mind is that,
for the polynomials \eqref{eq:pn_lim} to satisfy the orthogonality condition \eqref{eq:pmpn_ort_measure},
it is necessary that for every $\vphi$ in the weight function's support, $\vphi + \pi$ should also be included,
such that in the large-$n$, $m$ limit
\begin{equation}
    \frac{1}{2\pi} \int d\vphi\, w(\vphi) p_m(z)^* p_n(z)
    =
    \frac{1}{2\pi} \int d\vphi\, e^{i\vphi(n-m)}
    =
    \delta_{m n}
    \label{eq:pmpn_lim}
\end{equation}
holds.
Thus, this is a \textit{necessary condition for the system to attain the maximally ergodic regime},
and was implicitly assumed to hold in Eqs.~\eqref{eq:sequences_lim} and \eqref{eq:cn_t_lim}.
A particular and relevant instance in which such condition holds
is when the unitary and initial state chosen lead to a weight function with a support uniformly distributed on the unit circle,
which will become relevant in the discussion soon to follow.

Although the maximally ergodic regime presents itself in some different but equivalent ways,
we shall focus on only one of them to use as a signature of ergodicity.
In particular, the system's unitary in Krylov space $U_\mathcal{K}$
attains a distinct structure as it approaches the maximally ergodic regime.
Thus, we now define a measure of ergodicity through its normalized distance to a purely lower-diagonal matrix:
\begin{equation}
    \Erg(U)^{-1} = \frac{1}{\sqrt{2 \Tr(\mathbb{I})}} \norm{U_{\text{erg}} - U_\mathcal{K}}_{2,2}
    \label{eq:erg},
\end{equation}
where $U^{\text{erg}}_{n, m} = \delta_{n, m+1}$ and $\norm{\,\cdot\,}_{2,2}$ is the $L_{2,2}$ norm
(also known as the Frobenius norm).
The distance is normalized so that it lies in the interval $[0, 1]$,
since for any two unitaries $U$ and $V$ their distance is tightly bounded as
$\sqrt{2 \Tr(\mathbb{I})} \geq \norm{U - V}_{2,2} \geq 0$.
We define the ergodicity as the inverse of this distance,
so that this measure grows as the unitary approaches the maximally ergodic form in the Krylov basis.
This is not intended as a definitive measure of ergodicity,
but rather as a probe of the system's closeness to the maximally ergodic regime.
Our aim is then to follow this quantity's behavior in the integrability-to-chaos transition of various systems.
In that sense, the choice of this norm over others is somewhat arbitrary;
the relevant aspect is that the chosen norm is sensitive to the general structure of the unitary.
In Appendix~\ref{sec:norm_dependence} we show the main results of this work
compared with other choices of norm.
We point out as an observation that $U_{\text{erg}}$ is not a unitary matrix
since $\norm{U_\text{erg}}_{2,2}^2 = \Tr(\mathbb{I}) - 1$.
This implies that for physical unitary evolutions the exact maximally ergodic form
can only possibly be attained asymptotically for systems in the limit of large Hilbert space dimension.

Let us illustrate the points discussed so far with an example
highlighting that this framework applies also to autonomous unitary dynamics.
Consider a system with a time-independent Hamiltonian $H$,
for which we may write its unitary evolution as $U = e^{-i \tau H}$.
Since the Krylov space constructed by the Arnoldi iteration method involves only powers $U^t$ ($t\in\mathbb{N}_0$)
of the unitary, the time $\tau$ characterizes a minimal time step in the discretization of the time evolution.
The unitary's spectrum is that of $H$ wound around the unit circle,
yielding a uniform level density for a large enough value of $\tau$
compared with $\tau^* = \pi/\sigma_E$, where $\sigma_E^2$ is the Hamiltonian's spectral variance.
Thus, for time steps larger than this value the unitary will have support over the whole unit circle,
facilitating the emergence of the maximally ergodic regime, as previously discussed.
We show in Fig.~\ref{fig:SeqsUnifErg} (left panel) the ergodicity measure \eqref{eq:erg} in terms of the time step that is chosen to expand the unitary in Krylov space,
together with a quantity defined as
\begin{equation}
    \Delta_{\text{unif}}(U) = 2\,\underset{\vphi}{\text{sup}}\,|\CDF(\vphi) - \CDF_{\text{unif}}(\vphi)|
    \label{eq:uniformity}
\end{equation}
that measures how uniform the level density of the unitary is by calculating the distance of its cumulative distribution function (CDF) from that of a uniform distribution in the interval $[-\pi,\,\pi]$ \cite{CDFnote}.
In the right panel of Fig.~\ref{fig:SeqsUnifErg}
we show the Arnoldi sequences corresponding to the three values of $\tau$ highlighted in the left panel,
top to bottom respectively.
The data for this figure were obtained from the spin-chain model discussed in Sec.~\ref{sec:ising}
for $N=10$ with no disorder, $h_z=0.5$, and in its positive parity subspace (Hilbert space dimension of $528$).
The initial state $\psiz$ chosen to construct the Krylov space was with all spins in the ``down" configuration.
\begin{figure*}[ht]
    \centering
    \includegraphics[width=\linewidth]{SeqsUnifErg_IsingLT_10_hz_0.5_alldown.pdf}
    \caption{
    Left panel: inverse of the ergodicity measure (orange triangles) and measure of level density uniformity (blue diamonds) in terms of the chosen time step $\tau$, with $U = e^{-i\tau H}$ and $\sigma_E^2$ the Hamiltonian's spectral variance.
    Right panel: Arnoldi sequences arising from the time step values highlighted in the left panel (vertical dashed lines), top to bottom.
    }
    \label{fig:SeqsUnifErg}
\end{figure*}

In Fig.~\ref{fig:SeqsUnifErg} (left panel) we can see that, as $\tau$ increases, the unitary approaches the maximally ergodic form, but it saturates once its spectrum becomes uniform at a value of $\tau \sim \tau^*$. From a physical perspective this is not surprising: a large value of $\tau$ implies that in a single time step the system evolves further in time, and as a consequence less information is retained from the previous state. The time $\tau^*$ is also the typical timescale in which the survival probability $\abs{\braket{\psi_0}{\psi_\tau}}^2$ decays. For small $\tau$, the Arnoldi sequences (right panel) are consistent with continuous-time evolution, and they behave like Lanczos sequences for the Krylov expansion of $H$ \cite{nizami2024spreadcomplexityquantumchaos}. As the time step increases, larger portions of the sequences enter the maximally ergodic form.
We explore this issue further in Appendix~\ref{sec:timestep_dependence},
where we show our main results under different choices of time step.

These observations apply to both integrable and chaotic systems. However, if aspects (i) to (iii) as discussed in Sec.~\ref{sec:arnoldi_method} indeed define markers for quantum ergodicity, \textit{we would expect  chaotic systems to have a larger fraction of their Krylov space in the maximally ergodic regime
and the ergodicity measure \eqref{eq:erg} to saturate at a higher value}
provided that $\tau \gtrsim \tau^*$.
This motivates the comparison of the ergodicity measure defined in Eq.~\eqref{eq:erg} with known signatures of quantum chaos.

\section{Measures of quantum chaos}
\label{sec:chaos_measures}
The universal dynamical properties of quantum chaotic systems, such as their ergodic behavior,
are widely recognized as stemming from the statistical properties they exhibit akin to those of random matrices.
Specifically, the level-spacing and eigenvector statistics of chaotic systems can be described using specific ensembles of random matrices that correspond to the symmetries of the system~\cite{PhysRevLett.52.1, haake1991quantum, mehta2004random, wimberger2014nonlinear, PhysRevA.42.1013, MKus_1988, izrailev1987chaotic}. The models addressed in this work possess time-reversal symmetry, meaning the pertinent random matrices belong to the orthogonal ensemble. Therefore, we will assess the chaotic nature of these models by comparing both their level-spacing and eigenstate statistics with those of random matrices from the orthogonal ensemble.

As a measure of quantum chaos based on level-spacing statistics we utilize the mean value of the ratio of consecutive level spacings $s_i=e_i-e_{i-1}$ \cite{atas2013distribution},
\begin{equation}
    \langle \tilde r \rangle = \frac{1}{D} \sum_{n=1}^D \tilde r_n,
    \quad \text{where} \quad
    \tilde r_n = \frac{\min(s_n, s_{n-1})}{\max(s_n, s_{n-1})}
    \label{eq:r_tilde},
\end{equation}
which takes the values
$\langle \tilde r \rangle_{P} \approx 0.38629$
for Poissonnian statistics (integrable dynamics) and
$\langle \tilde r \rangle_{GOE} \approx 0.53590$ for the orthogonal ensemble (chaotic dynamics).
We normalize this quantity as
\begin{equation}
    \eta = \frac{\langle \tilde r \rangle - \langle \tilde r \rangle_{P}}{\langle \tilde r \rangle_{GOE} - \langle \tilde r \rangle_{P}}
    \label{eq:chaometer}
\end{equation}
to obtain a measure of quantum chaos such that
$\eta \approx 1$ for a chaotic system and $\eta \approx 0$ for an integrable system.

We complement this quantity with a chaotic measure based on eigenstate statistics.
By expanding each eigenstate of the system $\ket{\psi_i}$ onto those of
a suitable reference basis $\ket{\phi_j}$,
one obtains a set of coefficients $x \equiv \abs{c_{ij}}^2$
obeying a distribution that depends on the universality class of the system \cite{PhysRevA.42.1013, MKus_1988, izrailev1987chaotic}.
For the orthogonal ensemble this distribution is
\begin{equation}
    \PDF(x) = \frac{\Gamma(\frac{D}{2})}{\Gamma(\frac{D-1}{2} )} \frac{1}{\sqrt{\pi x}} (1 - x)^{\frac{D-3}{2}}
    \label{eq:evec_pdf},
\end{equation}
where $D$ is the dimension of the system and $\Gamma(z)$ the Gamma function.
We then define a measure of chaoticity using the Kolmogorov-Smirnov statistic~\cite{van2000asymptotic},
which compares the empirical CDF of the coefficient distribution to the CDF corresponding to the orthogonal ensemble distribution \eqref{eq:evec_pdf}:
\begin{equation}
    \KS = 1 - \underset{x}{\text{sup}}\,|\eCDF(x) - \CDF(x)|,
    \label{eq:KS}
\end{equation}
such that $\KS \approx 1$ for a chaotic system,
and decreases for integrable systems.

Eigenstate statistics depend on the chosen reference basis,
so this choice must be made with some care.
A reliable approach,
in the sense that the resulting statistics obey one of the universality classes if and only if the system in question is chaotic,
is to take a reference basis derived from an integrable system
\cite{doi:10.1080/00018732.2016.1198134, PhysRevA.34.591, PhysRevA.42.1013}.
In this work, we use the eigenbasis of the
Poissonian ensemble ($k=0$) for the random matrix model in Sec.~\ref{sec:rmt}, and the
interaction term $\sum_{i=1}^{N-1} \sigma_z^{(i)}\sigma_z^{(i+1)}$ for the spin chain in Secs.~\ref{sec:ising} and \ref{sec:ising_trotter}.
\section{Results}
\label{sec:results}
In this section, we present numerical results that compare the ergodicity
measure \eqref{eq:erg} with  spectral signatures of quantum chaos based on both
the statistics of level spacings and of the corresponding eigenstates (see
Sec.~\ref{sec:chaos_measures}). We achieve this by studying the parametric
integrability-to-chaos transition in the Hamiltonian evolution of a random
matrix model in Sec.~\ref{sec:rmt} and a spin-1/2 chain in
Sec.~\ref{sec:ising}. Additionally, in Sec.~\ref{sec:ising_trotter} we examine
the Trotterized unitary of this spin chain as we vary the same parameter.

In the first two cases the system evolves according to a time-independent Hamiltonian, from which we obtain the spectral statistics. On the other hand, for the Trotterized evolution we obtain the spectral statistics directly from the unitary, where the term level spacing refers to phase spacing. The generated Krylov space depends not only on the underlying unitary,
but also on the initial state chosen for its expansion.
In Ref.~\cite{PhysRevE.109.054209} it is shown that
in order to extract universal features from the structure of the Krylov space it is convenient to use initial states that are eigenstates of the system in its integrable regime.
This approach is employed in this study.
\subsection{The parametric RMT model}
\label{sec:rmt}

Random matrices were introduced in physics to describe complex
many-body quantum systems by Wigner~\cite{Wig55a}. Later on, it was conjectured
that such a description could arise in single-body settings, as long
as the system was chaotic~\cite{PhysRevLett.52.1,conjectureCasati}.
Such a conjecture has been widely tested~\cite{guhr98randomfull}
and is even taken as one of the definitions of quantum chaos~\cite{mehta2004random}.
Since then, it is customary to use random matrices to benchmark
chaotic dynamics, also called ergodic dynamics, and to compare quantitative
predictions of particular physical systems with predictions given by random matrix
ensembles, e.g., Ref.~\cite{pinedaseligmanELAF}.
Thus, the foundation for describing complexity in quantum mechanics lies
in the theory of random matrices. In fact, the eigenstate thermalization
hypothesis is intrinsically connected to this theory. Consequently, any
framework that aims to describe quantum complexity is expected to align with
the principles of RMT.
Depending on the physical situation, a given ensemble should be chosen to
compare. For example, for general situations, the Gaussian unitary ensemble
(GUE) is commonly used, whereas if the system to be described has a time-reversal
symmetry, the Gaussian orthogonal ensemble (GOE) is more precise, and for
describing a situation in which particles are singled out, other ensembles,
reflecting the tensor structure of the space are useful~\cite{Gorin_2008}.
In contrast to chaotic dynamics, integrable dynamics are often modeled with a
Poissonian spectrum.
Mixed dynamics, those that are not integrable but for
which chaos is not fully developed, can be modeled with an ensemble introduced
by Ref.~\cite{chavda2014poisson},
where the Hamiltonian is a superposition
of a diagonal Poissonian (to account for the integrable part)
and a GOE matrix (to account for the chaotic part),
with suitable weights.
Here, we opt for a variation on this model by replacing the GOE member by a banded random matrix.
Full random matrix models provide a useful basis
to analytically derive universal properties of quantum chaotic systems,
but they only describe local chaotic structures.
In contrast, banded random matrix models provide a more realistic description
of the statistical properties emerging from complex quantum systems \cite{CASATI1996430, Wig55a,kota2001embedded}.
Let
\begin{equation}
    H_{\text{RMT}} = \frac{H_0 + k V}{\sqrt{1 + k^2}}
    \label{eq:h_rmt},
\end{equation}
with $H_0$ a diagonal matrix with its nonzero elements chosen as real Gaussian
variables with unit standard deviation, describing the integrable part;
on the other hand, $V$
is taken to be a real-valued banded random matrix of bandwidth $2b$
(such that it is a full matrix for $b=D$, the system dimension)
whose values are Gaussian variables with standard deviation equal to $1/\sqrt{b+1}$,
such that its spectral variance is of order unity for large enough $b$.
In the spirit of Ref.~\cite{chavda2014poisson},
we define a transition parameter for this model as $\Lambda = k^2 D^2/2\pi(b + 1)$,
which makes the $\eta$ transition from Poissonian to GOE statistics
dimension-independent; see Appendix~\ref{sec:size_dependence}.

For this model we can directly compute $\Erg[{\rm exp}(-i \tau H_{\text{RMT}})]$
for some suitable time step $\tau$ and compare it with
the chaoticity measures $\eta$ and $\KS$.
In Fig.~\ref{fig:CvsE_randomfull}, we observe how
closely ergodicity tracks our selected measures of chaos.
For this figure, we have chosen $\tau = 1.5\,(2\pi)$
(so the spectrum has already spread over the unit circle),
a system dimension of $D=512$ and width $b=0.1D$.
The results are averaged over 10 realizations of the random Hamiltonian.
As an initial state, the center eigenstate of the Hamiltonian with $k=0$
(the integrable regime) was chosen.
In Appendix~\ref{sec:size_dependence} we show that this result
remains largely consistent upon changing the system dimension,
in the sense that the structure of Krylov space as measured by $\Erg$ reflects the chaotic transition.
\begin{figure}[ht] 
    \centering
    \includegraphics[width=\linewidth]{CvsE_RMTbanded_512_k_0_ev_256_tau_1.500_2pi_lambda.pdf}
    \caption{Chaotic measure $\eta$ (black filled circles) and $\KS$ (blue
filled squares) from the Hamiltonian
along with the unitary ergodicity measure $\Erg$ (orange triangles), as a function of the chaos
parameter $\Lambda$ across the statistical ensemble transition in the RMT model.
The $\KS$ and $\Erg$ curves in this figure were rescaled to the inverval $[0, 1]$ for ease of comparison.
}
    \label{fig:CvsE_randomfull}
\end{figure} 

\subsection{The spin chain}
\label{sec:ising}
Let us  consider a spin-1/2 chain  of length $N$ with open boundary conditions under a longitudinal-transversal magnetic field, described by the Hamiltonian
\begin{equation}
    H = \sum_{i=1}^N (\sigma_x^{(i)} + h_z \delta^{(i)} \sigma_z^{(i)})
        - \sum_{i=1}^{N-1} \sigma_z^{(i)}\sigma_z^{(i+1)}
    \label{eq:h_ising},
\end{equation}
where $\sigma_k^{(i)}$ is the Pauli operator at the $i$-th site in the direction $k$
and each $\delta^{(i)}$ is a disorder parameter drawn from a Gaussian distribution with unit mean and variance $\sigma^2$.
The purpose of introducing such inhomogeneities to the spin chain
is to remove values of $h_z$ for which the time step $\tau$ matches a periodicity of the time evolution.

When no disorder is present ($\delta^{(i)} = 1$ $\forall i$),
this model exhibits an integrability-to-chaos transition as a function of the strength of the longitudinal magnetic field $h_z$,
such that for $h_z = 0$ and $h_z \gg 1$ the system is integrable,
while for $h_z \approx 1$ it becomes chaotic.
This feature is still present upon the introduction of a weak enough amount of disorder. It is worth mentioning that this Hamiltonian has an antiunitary  symmetry reflected in the fact that in the computational basis, all matrix elements of Eq.~(\ref{eq:h_ising}) are real. Thus, we use the GOE ensemble for comparison.

We have computed the chaoticity of the system and the ergodicity from the unitary
for values of $h_z$ ranging from integrability to chaos.
Figure \ref{fig:CvsE_Isinghzdis_11_h0z=0,ev=1024} shows the ergodicity measure \eqref{eq:erg}
along with the chaotic measures \eqref{eq:chaometer} and \eqref{eq:KS},
highlighting the resemblance between the behavior of the ergodicity measure and that of the spectral measures of quantum chaos.
The curves in this figure were obtained by setting $N=11$ (Hilbert space dimension of $2^N=2048$),
a disorder strength of $\sigma = 0.4$, and by averaging over five disorder realizations.
The time step chosen to expand the unitary $U = e^{-i\tau H}$ on Krylov space
was $\tau = 0.15 \, (2\pi)$,
which ensures a uniform distribution of eigenphases on the unit circle for all values of $h_z$ here considered.
This figure was obtained considering the center eigenstate of $H$ in the integrable regime with $h_z=0$, as the initial state $\psiz$.
By choosing different eigenstates with $h_z=0$
or eigenstates from the integrable regime with $h_z=4$, we have observed entirely analogous results (data not shown).
\begin{figure}[ht]
    \centering
    \includegraphics[width=\linewidth]{CvsE_Isinghzdis_11_h0_hz_0_ev_1024_tau_0.15_2pi.pdf}
    \caption{Chaotic measures $\eta$ (black filled circles) and $\KS$ (blue filled squares) from the Hamiltonian
        along with the unitary ergodicity measure $\Erg$ (orange triangles) as a function of the chaos parameter $h_z$ across the integrability-to-chaos transition in the spin chain.
        The $\KS$ and $\Erg$ curves in this figure are rescaled to the interval $[0, 1]$ for ease of comparison.
    }
    \label{fig:CvsE_Isinghzdis_11_h0z=0,ev=1024}
\end{figure}

\subsection{The Trotterized spin chain}
\label{sec:ising_trotter}
So far we have computed the Krylov space and calculated the ergodicity measure
for conservative systems.
We now turn to a nonconservative system,
whose evolution is described by the unitary
\begin{eqnarray}
    U = e^{-i\tau H_{zz}} e^{-i\tau H_{s}}&,&
    \text{where } \,\,
    H_{zz} = - \sum_{i=1}^{N-1} \sigma_z^{(i)}\sigma_z^{(i+1)} \nonumber\\
    \quad \text{and} \quad
    H_{s} &=& \sum_{i=1}^N (\sigma_x^{(i)} + h_z \delta^{(i)} \sigma_z^{(i)})
    \label{eq:u_ising_trotter}.
\end{eqnarray}
This expression can be thought of as a first-order Trotter-Suzuki decomposition
for the evolution of the spin chain \eqref{eq:h_ising},
such that for small values of the Trotter step $\tau$
this unitary approximates the spin chain's evolution
\cite{trotter1959product, suzuki1976generalized, suzuki1985decomposition}.
For an arbitrary value of $\tau$,
Eq.~\eqref{eq:u_ising_trotter} is formally obtained as the evolution operator for the stroboscopic evolution
generated by the periodic time-dependent Hamiltonian
\begin{equation}
    H(t) = H_{zz} + \tau H_{s} \sum_{n=-\infty}^{\infty} \delta(t - n\tau)
    \label{eq:h_trotter},
\end{equation}
implying that energy generally is not conserved.
The specific Trotter-Suzuki decomposition \eqref{eq:u_ising_trotter} preserves the integrability-to-chaos transition from the original spin chain, although it may not produce the same evolution for large enough values of $\tau$.

In Fig.~\ref{fig:CvsE_t_Ising_11_alldown} we show results for the Trotterized spin chain,
again demonstrating that the ergodicity measure \eqref{eq:erg} is effective in signaling the chaotic transition in this non-autonomous system.
This is particularly interesting since this model exhibits a richer behavior as the chaotic parameter $h_z$ varies,
which stems from the emergence of regimes where the Trotter step $\tau$ matches a periodicity in the evolution.
\begin{figure}[ht]
    \centering
    \includegraphics[width=\linewidth]{CvsE_t_Isinghzdis_11_alldown_tau_0.6_2pi.pdf}
    \caption{Chaotic measures $\eta$ (black filled circles) and $\KS$ (blue filled squares) from the unitary
        along with its ergodicity measure $\Erg$ (orange triangles)
    as a function of the chaos parameter $h_z$ across the integrability-to-chaos transition in the Trotterized spin chain.
    The $\KS$ and $\Erg$ curves in this figure are rescaled to the interval $[0, 1]$ for ease of comparison.
}
    \label{fig:CvsE_t_Ising_11_alldown}
\end{figure}
This figure was obtained by setting $N=11$ (Hilbert space dimension of $2^N=2048$), a disorder strength of $\sigma = 0.1$, and by averaging over five disorder realizations. The Trotter step that we used is $\tau = 0.6\, (2\pi)$, which is large enough so that the evolution differs from that of the original spin chain \eqref{eq:h_ising},
and at the same time it yields a uniformly distributed phase spacing on the unit circle.
In this case, the unitary was expanded on Krylov space from an initial state with all spins in the
``down" configuration, which is an eigenstate of the system in the large-$h_z$ regime. Analogous results were obtained under different choices in the same regime (data not shown).

In this work we have focused on the structure of Krylov space through chaotic transitions of unitary dynamics under the lens of the maximally ergodic form.
Other signatures, such as Krylov complexity and the dispersion of Lanczos coefficients,
which were previously studied in the continuously time-evolved Krylov space obtained from the Lanczos approach
(see Refs.~\cite{parker2019, Rabinovici_2022_localization, Rabinovici_2022_integrability, PhysRevE.107.024217, PhysRevE.109.054209} and references therein)
can be easily extended to the Arnoldi approach.
This was recently explored in Ref.~\cite{nizami2024spreadcomplexityquantumchaos} in models related to the one in this section,
and they also found that these measures relate to the chaoticity in kicked systems,
complementing our results.

\section{Conclusions}
\label{sec:conclusions}

In this work, we study an ergodic measure that expands the Krylov complexity
formalism to encompass unitary circuit dynamics, with a particular emphasis on
Floquet circuits that emerge from the Trotter decomposition of Hamiltonian
dynamics. This completes the applicability of Krylov complexity to
non-autonomous systems, which are essential in new practical applications of
quantum technologies.  In the simulations of quantum many-body systems, the
identification of ergodic regimes can guide the development of more efficient
and accurate algorithms for quantum simulations. For example, in cases in which
memoryless dynamics (for instance in the thermalization process) or
randomized states are desired, one could allocate resources and reduce the
computational overhead.

More specifically, we aim to compare the new measure of quantum ergodicity
based on Krylov evolution with the conventional measures of quantum chaos
derived from spectral and eigenfunction statistics. This comparison was
conducted using random matrix models as well as unitary and Trotterized
evolutions of spin chains. We found that both measures similarly describe the
transition from integrability to chaos, reinforcing the idea that Krylov-based
quantum ergodicity is an appropriate tool for studying many-body systems.

There is another interesting implication of our findings. We have shown that
the ergodicity measure, as defined in Eq.(\ref{eq:erg}), effectively applies to
both autonomous and kicked systems, suggesting its versatility.  This
intriguing property motivates deeper exploration of its diverse characteristics
and behaviors.  Additionally, we observe a unifying principle across these
different measures: on one hand, they resemble spectral measures like the mean
value of the ratio of consecutive level spacings and eigenstate statistics,
while on the other hand, they also capture dynamic characteristics as
encapsulated in Eq.\ref{eq:erg}.

On the other hand, Krylov complexity appears to play a significant role in
quantum field theory. Several proposals have been put forward with the aim of
defining quantum complexity within the AdS/CFT dictionary. It is worth noting
that in Ref. \cite{rabinovici2023bulk} a precise correlation was established
between Krylov complexity in a one-dimensional quantum-mechanical boundary and a
specific bulk observable in two-dimensional quantum gravity theory.

Whereas previously the focus on this subject was
primarily on unitary circuit dynamics, here we also apply the same formalism to
autonomous Hamiltonian systems through its associated unitary evolution
operator, and we discuss the choice of time step necessary for the maximally
ergodic regime to be present.
For these types of systems there exists a “universal operator growth hypothesis” \cite{parker2019}.
We believe that our results constitute a first step toward understanding the connection between this and “maximally ergodic dynamics”.

\section{Acknowledgments}
This work has been partially supported by  CONICET (Grant No.~PIP
11220200100568CO), UBACyT (Grant  No.~ 20020220300049BA), ANPCyT
(PICT-2020-SERIEA-01082, PICT-2021-I-A-00654, PICT-2021-I-A-01288), and
UNAM-PAPIIT IG101324. DAW gratefully acknowledges support from ``Catedra Thomas
Brody'' (IF, UNAM).
\appendix
\section{Dependence on choice of norm}
We have defined a measure of ergodicity~\eqref{eq:erg} as the distance between
the system's unitary in Krylov space and its maximally ergodic form.  As there
is no unique way of defining such a distance, the ergodicity measure is not
uniquely defined.
In this appendix we show some of the results from the main
text under different choices of the norm used to define the system's unitary
distance to the maximally ergodic form.

The $L_{2, 2}$ norm (also known as the Frobenius norm) $\norm{A}_{2,2} =
\sqrt{\Trace(A^\dag A)}$ is used in the main text.
In Fig.~\ref{fig:RMT_Ising_norm_comp} we present the results for the parametric RMT model
and the Ising spin chain in Secs.~\ref{sec:rmt} and ~\ref{sec:ising},
respectively, with the addition of curves using the $L_{1, 1}$ norm
$\norm{A}_{1,1} = \sum_{ij} \abs{A_{ij}}$, the spectral 2-norm $\norm{A}_2 =
\sqrt{\lambda_{\text{max}}(A^\dagger A)}$, where $\lambda_\text{max}$
represents the largest eigenvalue, and the vector-induced 1-norm $\norm{A}_1 =
\max_{j}\sum_i \abs{A_{ij}}$.
In this figure all system parameters are the same as in the main text.
\label{sec:norm_dependence}
\begin{figure}[ht]
    \centering
    \includegraphics[width=\linewidth]{RMTbanded_ising_dis_norm_comp.pdf}
    \caption{Dependence with the choice of norm for the ergodicity measure $\Erg$
    compared with the chaotic measure $\eta$
    as a function of the chaos parameter through the integrability-to-chaos transition
    in (a) the RMT model and (b) the spin chain.
    The $\norm{\cdot}$ curves in this figure are rescaled to the inverval $[0,
1]$ for ease of comparison.
}
    \label{fig:RMT_Ising_norm_comp}
\end{figure}

These results show that the ergodic measure effectively reflects
the parametric chaotic transition when defined through the Euclidean norms $L_{2,2}$ and $L_2$.
Note that in Fig.~\ref{fig:RMT_Ising_norm_comp}.(b) the $L_{1,1}$ norm emphasizes the integrable regime $h_z \sim 0$,
where an accidental degeneracy occurs,
such that after rescaling to the interval $[0, 1]$ the rest of the curve appears to have a relatively large value.
Yet, in both cases this distance measure does track the transition.
The vector-induced 1-norm $\norm{\cdot}_1$ provides an example of an unsuitable choice,
since it is not sensitive to the general structure of the unitary:
information about ergodicity is mostly contained in its subdiagonal,
and this distance measure not only averages it out within each row,
but then extracts only a maximum, discarding the rest of the information.

\section{Dependence on time step}
\label{sec:timestep_dependence}
In Sec.~\ref{sec:ergodicity} we have argued that,
in order to attain the closest proximity to the maximally ergodic regime,
it is necessary for the time step associated with a single application of the unitary
to be greater than the typical survival probability decay time.
Figure~\ref{fig:SeqsUnifErg} shows the ergodicity measure~\eqref{eq:erg} as a function of the time step
as an example of this behavior.
In contrast, in this appendix we show the parametric chaotic transitions
for various choices of the time step in units of the characteristic decay time.

Figure~\ref{fig:RMT_Ising_erg_vs_time} shows the transitions
for both the RMT model in Sec.~\ref{sec:rmt} and the spin chain in Sec.~\ref{sec:ising},
for various time step choices.
Since in the Ising model the spectral variance is a function of the transition parameter $h_z$,
the time step is expressed in units of the \textit{mean} decay time
for the values of $h_z$ explored.
In this figure all system parameters, aside from the varying time step, are the same as in the main text.
\begin{figure}[ht]
    \centering
    \includegraphics[width=\linewidth]{RMTbanded_ising_dis_erg_vs_time.pdf}
    \caption{Dependence of the ergodicity measure $\Erg$
    as a function of the chaos parameter through the integrability-to-chaos transition in
    (a) the RMT model and (b) the spin chain,
    while varying the time step $\tau$ in units of the (mean) characteristic decay time.
    }
    \label{fig:RMT_Ising_erg_vs_time}
\end{figure}

Small time steps naturally lead to a small ergodicity,
almost independently of the transition parameter.
Interestingly, as the time step is increased to intermediate values in the order of the decay time,
the ergodicity measure does not straightforwardly follow the parametric transition.
In principle one might expect for it to do so,
although at lower values of ergodicity,
yet we see a peak and falloff for the RMT model,
and in the spin chain ergodicity increases up to the long time step curve, and follows it from thereon
(Figs.~\ref{fig:RMT_Ising_erg_vs_time}.(a) and (b), respectively).
Although interesting, a detailed analysis of this behavior is beyond the scope of this work.
Finally, for time steps larger than the decay time
the ergodicity does reach a convergence that coincides
with the parametric chaotic transition throughout,
as signaled by the spectral measures of quantum chaos as shown in the main text
(Figs.~\ref{fig:CvsE_randomfull} and ~\ref{fig:CvsE_Isinghzdis_11_h0z=0,ev=1024}).
\section{Dependence on system size}
\label{sec:size_dependence}
Quantum chaotic signatures typically present themselves in a robust manner
in the limit of large (enough) system sizes.
As commented in Sec.~\ref{sec:ergodicity},
the maximally ergodic regime is expected to be reached only asymptotically
for any physical unitary evolution.
In this appendix we explore how the ergodicity measure~\eqref{eq:erg}
varies in its ability to reflect the chaotic transition
in relation to the size of the system in question.

Figure~\ref{fig:RMT_Ising_erg_vs_size}
shows the ergodicity measure as a function of the transition parameter for various system sizes,
for both the RMT model in Sec.~\ref{sec:rmt} and the spin chain in Sec.~\ref{sec:ising}.
Save for the varying sizes,
all other system parameters to obtain these figures are the same as in the main text.
\begin{figure}[ht]
    \centering
    \includegraphics[width=\linewidth]{RMTbanded_ising_dis_erg_vs_size.pdf}
    \caption{Dependence of the ergodicity measure $\Erg$
    as a function of the chaos parameter through the integrability-to-chaos transition in
    (a) the RMT model and (b) the spin chain,
    while varying the system size.
    The curves in this figure are rescaled to the inverval $[0, 1]$ for ease of comparison.
}
    \label{fig:RMT_Ising_erg_vs_size}
\end{figure}

In both cases
the transition is reflected in the ergodicity measure
even at fairly modest system sizes, and
there is a clear convergence as it increases.
Notably, the ergodicity measure for the spin-chain model preserves its dependence
indicating that ergodic features can be found in this system at such sizes.
Such short spin chains have been found to possess quantum chaotic features
even in the extremely short limit~\cite{PhysRevE.103.L020201}.

Note that the ergodicity curves in Fig.~\ref{fig:RMT_Ising_erg_vs_size}
converge to a value that is \textit{smaller} as the dimension increases,
contrary to what one would expect from an ergodicity measure.
This is so only in appearance, due to the rescaling that is applied to each curve.
Without it, curves from a system with higher Hilbert space dimension yield a higher overall ergodicity
(provided the system is in a chaotic or mixed regime).
\bibliography{referencias}

\begin{thebibliography}{49}%
\makeatletter
\providecommand \@ifxundefined [1]{%
 \@ifx{#1\undefined}
}%
\providecommand \@ifnum [1]{%
 \ifnum #1\expandafter \@firstoftwo
 \else \expandafter \@secondoftwo
 \fi
}%
\providecommand \@ifx [1]{%
 \ifx #1\expandafter \@firstoftwo
 \else \expandafter \@secondoftwo
 \fi
}%
\providecommand \natexlab [1]{#1}%
\providecommand \enquote  [1]{``#1''}%
\providecommand \bibnamefont  [1]{#1}%
\providecommand \bibfnamefont [1]{#1}%
\providecommand \citenamefont [1]{#1}%
\providecommand \href@noop [0]{\@secondoftwo}%
\providecommand \href [0]{\begingroup \@sanitize@url \@href}%
\providecommand \@href[1]{\@@startlink{#1}\@@href}%
\providecommand \@@href[1]{\endgroup#1\@@endlink}%
\providecommand \@sanitize@url [0]{\catcode `\\12\catcode `\$12\catcode `\&12\catcode `\#12\catcode `\^12\catcode `\_12\catcode `\%12\relax}%
\providecommand \@@startlink[1]{}%
\providecommand \@@endlink[0]{}%
\providecommand \url  [0]{\begingroup\@sanitize@url \@url }%
\providecommand \@url [1]{\endgroup\@href {#1}{\urlprefix }}%
\providecommand \urlprefix  [0]{URL }%
\providecommand \Eprint [0]{\href }%
\providecommand \doibase [0]{https://doi.org/}%
\providecommand \selectlanguage [0]{\@gobble}%
\providecommand \bibinfo  [0]{\@secondoftwo}%
\providecommand \bibfield  [0]{\@secondoftwo}%
\providecommand \translation [1]{[#1]}%
\providecommand \BibitemOpen [0]{}%
\providecommand \bibitemStop [0]{}%
\providecommand \bibitemNoStop [0]{.\EOS\space}%
\providecommand \EOS [0]{\spacefactor3000\relax}%
\providecommand \BibitemShut  [1]{\csname bibitem#1\endcsname}%
\let\auto@bib@innerbib\@empty
\bibitem [{\citenamefont {Schweigler}\ \emph {et~al.}(2017)\citenamefont {Schweigler}, \citenamefont {Kasper}, \citenamefont {Erne}, \citenamefont {Mazets}, \citenamefont {Rauer}, \citenamefont {Cataldini}, \citenamefont {Langen}, \citenamefont {Gasenzer}, \citenamefont {Berges},\ and\ \citenamefont {Schmiedmayer}}]{schweigler2017experimental}%
  \BibitemOpen
  \bibfield  {author} {\bibinfo {author} {\bibfnamefont {T.}~\bibnamefont {Schweigler}}, \bibinfo {author} {\bibfnamefont {V.}~\bibnamefont {Kasper}}, \bibinfo {author} {\bibfnamefont {S.}~\bibnamefont {Erne}}, \bibinfo {author} {\bibfnamefont {I.}~\bibnamefont {Mazets}}, \bibinfo {author} {\bibfnamefont {B.}~\bibnamefont {Rauer}}, \bibinfo {author} {\bibfnamefont {F.}~\bibnamefont {Cataldini}}, \bibinfo {author} {\bibfnamefont {T.}~\bibnamefont {Langen}}, \bibinfo {author} {\bibfnamefont {T.}~\bibnamefont {Gasenzer}}, \bibinfo {author} {\bibfnamefont {J.}~\bibnamefont {Berges}},\ and\ \bibinfo {author} {\bibfnamefont {J.}~\bibnamefont {Schmiedmayer}},\ }\bibfield  {title} {\bibinfo {title} {Experimental characterization of a quantum many-body system via higher-order correlations},\ }\href@noop {} {\bibfield  {journal} {\bibinfo  {journal} {Nature}\ }\textbf {\bibinfo {volume} {545}},\ \bibinfo {pages} {323} (\bibinfo {year} {2017})}\BibitemShut {NoStop}%
\bibitem [{\citenamefont {Zhang}\ \emph {et~al.}(2017)\citenamefont {Zhang}, \citenamefont {Pagano}, \citenamefont {Hess}, \citenamefont {Kyprianidis}, \citenamefont {Becker}, \citenamefont {Kaplan}, \citenamefont {Gorshkov}, \citenamefont {Gong},\ and\ \citenamefont {Monroe}}]{zhang2017observation}%
  \BibitemOpen
  \bibfield  {author} {\bibinfo {author} {\bibfnamefont {J.}~\bibnamefont {Zhang}}, \bibinfo {author} {\bibfnamefont {G.}~\bibnamefont {Pagano}}, \bibinfo {author} {\bibfnamefont {P.~W.}\ \bibnamefont {Hess}}, \bibinfo {author} {\bibfnamefont {A.}~\bibnamefont {Kyprianidis}}, \bibinfo {author} {\bibfnamefont {P.}~\bibnamefont {Becker}}, \bibinfo {author} {\bibfnamefont {H.}~\bibnamefont {Kaplan}}, \bibinfo {author} {\bibfnamefont {A.~V.}\ \bibnamefont {Gorshkov}}, \bibinfo {author} {\bibfnamefont {Z.-X.}\ \bibnamefont {Gong}},\ and\ \bibinfo {author} {\bibfnamefont {C.}~\bibnamefont {Monroe}},\ }\bibfield  {title} {\bibinfo {title} {Observation of a many-body dynamical phase transition with a 53-qubit quantum simulator},\ }\href@noop {} {\bibfield  {journal} {\bibinfo  {journal} {Nature}\ }\textbf {\bibinfo {volume} {551}},\ \bibinfo {pages} {601} (\bibinfo {year} {2017})}\BibitemShut {NoStop}%
\bibitem [{\citenamefont {Parker}\ \emph {et~al.}(2019)\citenamefont {Parker}, \citenamefont {Cao}, \citenamefont {Avdoshkin}, \citenamefont {Scaffidi},\ and\ \citenamefont {Altman}}]{parker2019}%
  \BibitemOpen
  \bibfield  {author} {\bibinfo {author} {\bibfnamefont {D.~E.}\ \bibnamefont {Parker}}, \bibinfo {author} {\bibfnamefont {X.}~\bibnamefont {Cao}}, \bibinfo {author} {\bibfnamefont {A.}~\bibnamefont {Avdoshkin}}, \bibinfo {author} {\bibfnamefont {T.}~\bibnamefont {Scaffidi}},\ and\ \bibinfo {author} {\bibfnamefont {E.}~\bibnamefont {Altman}},\ }\bibfield  {title} {\bibinfo {title} {A universal operator growth hypothesis},\ }\href {https://doi.org/10.1103/PhysRevX.9.041017} {\bibfield  {journal} {\bibinfo  {journal} {Phys. Rev. X}\ }\textbf {\bibinfo {volume} {9}},\ \bibinfo {pages} {041017} (\bibinfo {year} {2019})}\BibitemShut {NoStop}%
\bibitem [{\citenamefont {Rabinovici}\ \emph {et~al.}(2021)\citenamefont {Rabinovici}, \citenamefont {S{\'{a}}nchez-Garrido}, \citenamefont {Shir},\ and\ \citenamefont {Sonner}}]{Rabinovici2021}%
  \BibitemOpen
  \bibfield  {author} {\bibinfo {author} {\bibfnamefont {E.}~\bibnamefont {Rabinovici}}, \bibinfo {author} {\bibfnamefont {A.}~\bibnamefont {S{\'{a}}nchez-Garrido}}, \bibinfo {author} {\bibfnamefont {R.}~\bibnamefont {Shir}},\ and\ \bibinfo {author} {\bibfnamefont {J.}~\bibnamefont {Sonner}},\ }\bibfield  {title} {\bibinfo {title} {Operator complexity: a journey to the edge of krylov space},\ }\href {https://doi.org/10.1007/jhep06(2021)062} {\bibfield  {journal} {\bibinfo  {journal} {J. High Energy Phys.}\ }\textbf {\bibinfo {volume} {2021}}\bibinfo  {number} { (6)}}\BibitemShut {NoStop}%
\bibitem [{\citenamefont {Rabinovici}\ \emph {et~al.}(2022{\natexlab{a}})\citenamefont {Rabinovici}, \citenamefont {S{\'{a}}nchez-Garrido}, \citenamefont {Shir},\ and\ \citenamefont {Sonner}}]{Rabinovici_2022_integrability}%
  \BibitemOpen
\bibfield  {number} {  }\bibfield  {author} {\bibinfo {author} {\bibfnamefont {E.}~\bibnamefont {Rabinovici}}, \bibinfo {author} {\bibfnamefont {A.}~\bibnamefont {S{\'{a}}nchez-Garrido}}, \bibinfo {author} {\bibfnamefont {R.}~\bibnamefont {Shir}},\ and\ \bibinfo {author} {\bibfnamefont {J.}~\bibnamefont {Sonner}},\ }\bibfield  {title} {\bibinfo {title} {Krylov complexity from integrability to chaos},\ }\bibfield  {journal} {\bibinfo  {journal} {Journal of High Energy Physics}\ }\textbf {\bibinfo {volume} {2022}},\ \href {https://doi.org/10.1007/jhep07(2022)151} {10.1007/jhep07(2022)151} (\bibinfo {year} {2022}{\natexlab{a}})\BibitemShut {NoStop}%
\bibitem [{\citenamefont {Balasubramanian}\ \emph {et~al.}(2022)\citenamefont {Balasubramanian}, \citenamefont {Caputa}, \citenamefont {Magan},\ and\ \citenamefont {Wu}}]{Balasubramanian_2022}%
  \BibitemOpen
  \bibfield  {author} {\bibinfo {author} {\bibfnamefont {V.}~\bibnamefont {Balasubramanian}}, \bibinfo {author} {\bibfnamefont {P.}~\bibnamefont {Caputa}}, \bibinfo {author} {\bibfnamefont {J.~M.}\ \bibnamefont {Magan}},\ and\ \bibinfo {author} {\bibfnamefont {Q.}~\bibnamefont {Wu}},\ }\bibfield  {title} {\bibinfo {title} {Quantum chaos and the complexity of spread of states},\ }\href {https://doi.org/10.1103/PhysRevD.106.046007} {\bibfield  {journal} {\bibinfo  {journal} {Phys. Rev. D}\ }\textbf {\bibinfo {volume} {106}},\ \bibinfo {pages} {046007} (\bibinfo {year} {2022})}\BibitemShut {NoStop}%
\bibitem [{\citenamefont {Espa\~nol}\ and\ \citenamefont {Wisniacki}(2023)}]{PhysRevE.107.024217}%
  \BibitemOpen
  \bibfield  {author} {\bibinfo {author} {\bibfnamefont {B.~L.}\ \bibnamefont {Espa\~nol}}\ and\ \bibinfo {author} {\bibfnamefont {D.~A.}\ \bibnamefont {Wisniacki}},\ }\bibfield  {title} {\bibinfo {title} {Assessing the saturation of krylov complexity as a measure of chaos},\ }\href {https://doi.org/10.1103/PhysRevE.107.024217} {\bibfield  {journal} {\bibinfo  {journal} {Phys. Rev. E}\ }\textbf {\bibinfo {volume} {107}},\ \bibinfo {pages} {024217} (\bibinfo {year} {2023})}\BibitemShut {NoStop}%
\bibitem [{\citenamefont {Scialchi}\ \emph {et~al.}(2024)\citenamefont {Scialchi}, \citenamefont {Roncaglia},\ and\ \citenamefont {Wisniacki}}]{PhysRevE.109.054209}%
  \BibitemOpen
  \bibfield  {author} {\bibinfo {author} {\bibfnamefont {G.~F.}\ \bibnamefont {Scialchi}}, \bibinfo {author} {\bibfnamefont {A.~J.}\ \bibnamefont {Roncaglia}},\ and\ \bibinfo {author} {\bibfnamefont {D.~A.}\ \bibnamefont {Wisniacki}},\ }\bibfield  {title} {\bibinfo {title} {Integrability-to-chaos transition through the krylov approach for state evolution},\ }\href {https://doi.org/10.1103/PhysRevE.109.054209} {\bibfield  {journal} {\bibinfo  {journal} {Phys. Rev. E}\ }\textbf {\bibinfo {volume} {109}},\ \bibinfo {pages} {054209} (\bibinfo {year} {2024})}\BibitemShut {NoStop}%
\bibitem [{\citenamefont {Domingo}\ \emph {et~al.}(2023)\citenamefont {Domingo}, \citenamefont {Borondo}, \citenamefont {Scialchi}, \citenamefont {Roncaglia}, \citenamefont {Carlo},\ and\ \citenamefont {Wisniacki}}]{domingo2023quantum}%
  \BibitemOpen
  \bibfield  {author} {\bibinfo {author} {\bibfnamefont {L.}~\bibnamefont {Domingo}}, \bibinfo {author} {\bibfnamefont {F.}~\bibnamefont {Borondo}}, \bibinfo {author} {\bibfnamefont {G.}~\bibnamefont {Scialchi}}, \bibinfo {author} {\bibfnamefont {A.~J.}\ \bibnamefont {Roncaglia}}, \bibinfo {author} {\bibfnamefont {G.~G.}\ \bibnamefont {Carlo}},\ and\ \bibinfo {author} {\bibfnamefont {D.~A.}\ \bibnamefont {Wisniacki}},\ }\href@noop {} {\bibinfo {title} {Quantum reservoir complexity by krylov evolution approach}} (\bibinfo {year} {2023}),\ \Eprint {https://arxiv.org/abs/2310.00790} {arXiv:2310.00790 [quant-ph]} \BibitemShut {NoStop}%
\bibitem [{\citenamefont {Nizami}\ and\ \citenamefont {Shrestha}(2023)}]{nizami2023krylov}%
  \BibitemOpen
  \bibfield  {author} {\bibinfo {author} {\bibfnamefont {A.~A.}\ \bibnamefont {Nizami}}\ and\ \bibinfo {author} {\bibfnamefont {A.~W.}\ \bibnamefont {Shrestha}},\ }\bibfield  {title} {\bibinfo {title} {Krylov construction and complexity for driven quantum systems},\ }\href@noop {} {\bibfield  {journal} {\bibinfo  {journal} {Physical Review E}\ }\textbf {\bibinfo {volume} {108}},\ \bibinfo {pages} {054222} (\bibinfo {year} {2023})}\BibitemShut {NoStop}%
\bibitem [{\citenamefont {Suchsland}\ \emph {et~al.}(2025)\citenamefont {Suchsland}, \citenamefont {Moessner},\ and\ \citenamefont {Claeys}}]{suchsland2023krylov}%
  \BibitemOpen
  \bibfield  {author} {\bibinfo {author} {\bibfnamefont {P.}~\bibnamefont {Suchsland}}, \bibinfo {author} {\bibfnamefont {R.}~\bibnamefont {Moessner}},\ and\ \bibinfo {author} {\bibfnamefont {P.~W.}\ \bibnamefont {Claeys}},\ }\bibfield  {title} {\bibinfo {title} {Krylov complexity and trotter transitions in unitary circuit dynamics},\ }\href {https://doi.org/10.1103/PhysRevB.111.014309} {\bibfield  {journal} {\bibinfo  {journal} {Phys. Rev. B}\ }\textbf {\bibinfo {volume} {111}},\ \bibinfo {pages} {014309} (\bibinfo {year} {2025})}\BibitemShut {NoStop}%
\bibitem [{\citenamefont {Nizami}\ and\ \citenamefont {Shrestha}(2024)}]{nizami2024spreadcomplexityquantumchaos}%
  \BibitemOpen
  \bibfield  {author} {\bibinfo {author} {\bibfnamefont {A.~A.}\ \bibnamefont {Nizami}}\ and\ \bibinfo {author} {\bibfnamefont {A.~W.}\ \bibnamefont {Shrestha}},\ }\bibfield  {title} {\bibinfo {title} {Spread complexity and quantum chaos for periodically driven spin-chains},\ }\href {https://arxiv.org/abs/2405.16182} {\bibfield  {journal} {\bibinfo  {journal} {arXiv}\ } (\bibinfo {year} {2024})},\ \Eprint {https://arxiv.org/abs/2405.16182} {arXiv:2405.16182 [quant-ph]} \BibitemShut {NoStop}%
\bibitem [{\citenamefont {Daley}\ \emph {et~al.}(2022)\citenamefont {Daley}, \citenamefont {Bloch}, \citenamefont {Kokail}, \citenamefont {Flannigan}, \citenamefont {Pearson}, \citenamefont {Troyer},\ and\ \citenamefont {Zoller}}]{daley2022practical}%
  \BibitemOpen
  \bibfield  {author} {\bibinfo {author} {\bibfnamefont {A.~J.}\ \bibnamefont {Daley}}, \bibinfo {author} {\bibfnamefont {I.}~\bibnamefont {Bloch}}, \bibinfo {author} {\bibfnamefont {C.}~\bibnamefont {Kokail}}, \bibinfo {author} {\bibfnamefont {S.}~\bibnamefont {Flannigan}}, \bibinfo {author} {\bibfnamefont {N.}~\bibnamefont {Pearson}}, \bibinfo {author} {\bibfnamefont {M.}~\bibnamefont {Troyer}},\ and\ \bibinfo {author} {\bibfnamefont {P.}~\bibnamefont {Zoller}},\ }\bibfield  {title} {\bibinfo {title} {Practical quantum advantage in quantum simulation},\ }\href@noop {} {\bibfield  {journal} {\bibinfo  {journal} {Nature}\ }\textbf {\bibinfo {volume} {607}},\ \bibinfo {pages} {667} (\bibinfo {year} {2022})}\BibitemShut {NoStop}%
\bibitem [{\citenamefont {Berthusen}\ \emph {et~al.}(2022)\citenamefont {Berthusen}, \citenamefont {Trevisan}, \citenamefont {Iadecola},\ and\ \citenamefont {Orth}}]{berthusen2022quantum}%
  \BibitemOpen
  \bibfield  {author} {\bibinfo {author} {\bibfnamefont {N.~F.}\ \bibnamefont {Berthusen}}, \bibinfo {author} {\bibfnamefont {T.~V.}\ \bibnamefont {Trevisan}}, \bibinfo {author} {\bibfnamefont {T.}~\bibnamefont {Iadecola}},\ and\ \bibinfo {author} {\bibfnamefont {P.~P.}\ \bibnamefont {Orth}},\ }\bibfield  {title} {\bibinfo {title} {Quantum dynamics simulations beyond the coherence time on noisy intermediate-scale quantum hardware by variational trotter compression},\ }\href@noop {} {\bibfield  {journal} {\bibinfo  {journal} {Physical Review Research}\ }\textbf {\bibinfo {volume} {4}},\ \bibinfo {pages} {023097} (\bibinfo {year} {2022})}\BibitemShut {NoStop}%
\bibitem [{\citenamefont {Arnoldi}(1951)}]{arnoldi1951principle}%
  \BibitemOpen
  \bibfield  {author} {\bibinfo {author} {\bibfnamefont {W.~E.}\ \bibnamefont {Arnoldi}},\ }\bibfield  {title} {\bibinfo {title} {The principle of minimized iterations in the solution of the matrix eigenvalue problem},\ }\href@noop {} {\bibfield  {journal} {\bibinfo  {journal} {Quarterly of applied mathematics}\ }\textbf {\bibinfo {volume} {9}},\ \bibinfo {pages} {17} (\bibinfo {year} {1951})}\BibitemShut {NoStop}%
\bibitem [{\citenamefont {Bai}\ and\ \citenamefont {Su}(2005)}]{bai2005soar}%
  \BibitemOpen
  \bibfield  {author} {\bibinfo {author} {\bibfnamefont {Z.}~\bibnamefont {Bai}}\ and\ \bibinfo {author} {\bibfnamefont {Y.}~\bibnamefont {Su}},\ }\bibfield  {title} {\bibinfo {title} {Soar: A second-order arnoldi method for the solution of the quadratic eigenvalue problem},\ }\href@noop {} {\bibfield  {journal} {\bibinfo  {journal} {SIAM Journal on Matrix Analysis and Applications}\ }\textbf {\bibinfo {volume} {26}},\ \bibinfo {pages} {640} (\bibinfo {year} {2005})}\BibitemShut {NoStop}%
\bibitem [{\citenamefont {Bellalij}\ \emph {et~al.}(2007)\citenamefont {Bellalij}, \citenamefont {Saad},\ and\ \citenamefont {Sadok}}]{bellalij2007convergence}%
  \BibitemOpen
  \bibfield  {author} {\bibinfo {author} {\bibfnamefont {M.}~\bibnamefont {Bellalij}}, \bibinfo {author} {\bibfnamefont {Y.}~\bibnamefont {Saad}},\ and\ \bibinfo {author} {\bibfnamefont {H.}~\bibnamefont {Sadok}},\ }\bibfield  {title} {\bibinfo {title} {On the convergence of the arnoldi process for eigenvalue problems, report umsi-2007-12, minnesota supercomputer institute},\ }\href@noop {} {\bibfield  {journal} {\bibinfo  {journal} {University of Minnesota, Minneapolis, MN}\ } (\bibinfo {year} {2007})}\BibitemShut {NoStop}%
\bibitem [{\citenamefont {Bhattacharya}\ \emph {et~al.}(2022)\citenamefont {Bhattacharya}, \citenamefont {Nandy}, \citenamefont {Nath},\ and\ \citenamefont {Sahu}}]{Bhattacharya:2022gbz}%
  \BibitemOpen
  \bibfield  {author} {\bibinfo {author} {\bibfnamefont {A.}~\bibnamefont {Bhattacharya}}, \bibinfo {author} {\bibfnamefont {P.}~\bibnamefont {Nandy}}, \bibinfo {author} {\bibfnamefont {P.~P.}\ \bibnamefont {Nath}},\ and\ \bibinfo {author} {\bibfnamefont {H.}~\bibnamefont {Sahu}},\ }\bibfield  {title} {\bibinfo {title} {{Operator growth and Krylov construction in dissipative open quantum systems}},\ }\href {https://doi.org/10.1007/JHEP12(2022)081} {\bibfield  {journal} {\bibinfo  {journal} {JHEP}\ }\textbf {\bibinfo {volume} {12}},\ \bibinfo {pages} {081}},\ \Eprint {https://arxiv.org/abs/2207.05347} {arXiv:2207.05347 [quant-ph]} \BibitemShut {NoStop}%
\bibitem [{\citenamefont {Notenson}\ \emph {et~al.}(2023)\citenamefont {Notenson}, \citenamefont {Garc{\'\i}a-Mata}, \citenamefont {Roncaglia},\ and\ \citenamefont {Wisniacki}}]{Garcia-Mata:2023}%
  \BibitemOpen
  \bibfield  {author} {\bibinfo {author} {\bibfnamefont {T.}~\bibnamefont {Notenson}}, \bibinfo {author} {\bibfnamefont {I.}~\bibnamefont {Garc{\'\i}a-Mata}}, \bibinfo {author} {\bibfnamefont {A.~J.}\ \bibnamefont {Roncaglia}},\ and\ \bibinfo {author} {\bibfnamefont {D.~A.}\ \bibnamefont {Wisniacki}},\ }\bibfield  {title} {\bibinfo {title} {Classical approach to equilibrium of out-of-time ordered correlators in mixed systems},\ }\href@noop {} {\bibfield  {journal} {\bibinfo  {journal} {Physical Review E}\ }\textbf {\bibinfo {volume} {107}},\ \bibinfo {pages} {064207} (\bibinfo {year} {2023})}\BibitemShut {NoStop}%
\bibitem [{\citenamefont {Xu}\ and\ \citenamefont {Swingle}(2024)}]{PRXQuantum.5.010201}%
  \BibitemOpen
  \bibfield  {author} {\bibinfo {author} {\bibfnamefont {S.}~\bibnamefont {Xu}}\ and\ \bibinfo {author} {\bibfnamefont {B.}~\bibnamefont {Swingle}},\ }\bibfield  {title} {\bibinfo {title} {Scrambling dynamics and out-of-time-ordered correlators in quantum many-body systems},\ }\href {https://doi.org/10.1103/PRXQuantum.5.010201} {\bibfield  {journal} {\bibinfo  {journal} {PRX Quantum}\ }\textbf {\bibinfo {volume} {5}},\ \bibinfo {pages} {010201} (\bibinfo {year} {2024})}\BibitemShut {NoStop}%
\bibitem [{\citenamefont {Swingle}(2018)}]{swingle2018unscrambling}%
  \BibitemOpen
  \bibfield  {author} {\bibinfo {author} {\bibfnamefont {B.}~\bibnamefont {Swingle}},\ }\bibfield  {title} {\bibinfo {title} {Unscrambling the physics of out-of-time-order correlators},\ }\href@noop {} {\bibfield  {journal} {\bibinfo  {journal} {Nature Physics}\ }\textbf {\bibinfo {volume} {14}},\ \bibinfo {pages} {988} (\bibinfo {year} {2018})}\BibitemShut {NoStop}%
\bibitem [{\citenamefont {Nandy}\ \emph {et~al.}(2024)\citenamefont {Nandy}, \citenamefont {Matsoukas-Roubeas}, \citenamefont {Mart{\'\i}nez-Azcona}, \citenamefont {Dymarsky},\ and\ \citenamefont {del Campo}}]{nandy2024quantum}%
  \BibitemOpen
  \bibfield  {author} {\bibinfo {author} {\bibfnamefont {P.}~\bibnamefont {Nandy}}, \bibinfo {author} {\bibfnamefont {A.~S.}\ \bibnamefont {Matsoukas-Roubeas}}, \bibinfo {author} {\bibfnamefont {P.}~\bibnamefont {Mart{\'\i}nez-Azcona}}, \bibinfo {author} {\bibfnamefont {A.}~\bibnamefont {Dymarsky}},\ and\ \bibinfo {author} {\bibfnamefont {A.}~\bibnamefont {del Campo}},\ }\bibfield  {title} {\bibinfo {title} {Quantum dynamics in krylov space: Methods and applications},\ }\href@noop {} {\bibfield  {journal} {\bibinfo  {journal} {arXiv preprint arXiv:2405.09628}\ } (\bibinfo {year} {2024})}\BibitemShut {NoStop}%
\bibitem [{\citenamefont {Szego}(1939)}]{szego}%
  \BibitemOpen
  \bibfield  {author} {\bibinfo {author} {\bibfnamefont {G.}~\bibnamefont {Szego}},\ }\href@noop {} {\emph {\bibinfo {title} {Orthogonal polynomials}}},\ Vol.~\bibinfo {volume} {23}\ (\bibinfo  {publisher} {American Mathematical Society, Providence, Rhode Island},\ \bibinfo {year} {1939})\ p.\ \bibinfo {pages} {297},\ \bibinfo {note} {see also p. 276}\BibitemShut {NoStop}%
\bibitem [{CDF()}]{CDFnote}%
  \BibitemOpen
  \href@noop {} {}\bibinfo {note} {Let $\rho_l(\varphi)$ be the unitary's level density, its cumulative distribution function (CDF) is defined by $\CDF(\varphi) = \int_{-\pi}^\varphi d\theta\, \rho_l(\theta)$. For a uniform distribution $\rho_{\text{unif}}(\varphi) = \frac{1}{2\pi}$ the corresponding CDF is $\CDF_{\text{unif}}(\varphi) = \frac{\varphi + \pi}{2\pi}$. The quantity $\Delta{\text{unif}}(U)$ can then be understood as a Kolmovogorv-Smirnov statistic quantifying how close the level density is to being uniform \cite{van2000asymptotic}.}\BibitemShut {Stop}%
\bibitem [{\citenamefont {Bohigas}\ \emph {et~al.}(1984)\citenamefont {Bohigas}, \citenamefont {Giannoni},\ and\ \citenamefont {Schmit}}]{PhysRevLett.52.1}%
  \BibitemOpen
  \bibfield  {author} {\bibinfo {author} {\bibfnamefont {O.}~\bibnamefont {Bohigas}}, \bibinfo {author} {\bibfnamefont {M.~J.}\ \bibnamefont {Giannoni}},\ and\ \bibinfo {author} {\bibfnamefont {C.}~\bibnamefont {Schmit}},\ }\bibfield  {title} {\bibinfo {title} {Characterization of chaotic quantum spectra and universality of level fluctuation laws},\ }\href {https://doi.org/10.1103/PhysRevLett.52.1} {\bibfield  {journal} {\bibinfo  {journal} {Phys. Rev. Lett.}\ }\textbf {\bibinfo {volume} {52}},\ \bibinfo {pages} {1} (\bibinfo {year} {1984})}\BibitemShut {NoStop}%
\bibitem [{\citenamefont {Haake}(1991)}]{haake1991quantum}%
  \BibitemOpen
  \bibfield  {author} {\bibinfo {author} {\bibfnamefont {F.}~\bibnamefont {Haake}},\ }\href@noop {} {\emph {\bibinfo {title} {Quantum signatures of chaos}}}\ (\bibinfo  {publisher} {Springer, New York},\ \bibinfo {year} {1991})\BibitemShut {NoStop}%
\bibitem [{\citenamefont {Mehta}(2004)}]{mehta2004random}%
  \BibitemOpen
  \bibfield  {author} {\bibinfo {author} {\bibfnamefont {M.~L.}\ \bibnamefont {Mehta}},\ }\href@noop {} {\emph {\bibinfo {title} {Random matrices}}}\ (\bibinfo  {publisher} {Elsevier},\ \bibinfo {year} {2004})\BibitemShut {NoStop}%
\bibitem [{\citenamefont {Wimberger}(2014)}]{wimberger2014nonlinear}%
  \BibitemOpen
  \bibfield  {author} {\bibinfo {author} {\bibfnamefont {S.}~\bibnamefont {Wimberger}},\ }\href@noop {} {\emph {\bibinfo {title} {Nonlinear dynamics and quantum chaos}}},\ Vol.~\bibinfo {volume} {10}\ (\bibinfo  {publisher} {Springer, Cham, Switzerland},\ \bibinfo {year} {2014})\BibitemShut {NoStop}%
\bibitem [{\citenamefont {Haake}\ and\ \citenamefont {\ifmmode~\dot{Z}\else \.{Z}\fi{}yczkowski}(1990)}]{PhysRevA.42.1013}%
  \BibitemOpen
  \bibfield  {author} {\bibinfo {author} {\bibfnamefont {F.}~\bibnamefont {Haake}}\ and\ \bibinfo {author} {\bibfnamefont {K.}~\bibnamefont {\ifmmode~\dot{Z}\else \.{Z}\fi{}yczkowski}},\ }\bibfield  {title} {\bibinfo {title} {Random-matrix theory and eigenmodes of dynamical systems},\ }\href {https://doi.org/10.1103/PhysRevA.42.1013} {\bibfield  {journal} {\bibinfo  {journal} {Phys. Rev. A}\ }\textbf {\bibinfo {volume} {42}},\ \bibinfo {pages} {1013} (\bibinfo {year} {1990})}\BibitemShut {NoStop}%
\bibitem [{\citenamefont {Kus}\ \emph {et~al.}(1988)\citenamefont {Kus}, \citenamefont {Mostowski},\ and\ \citenamefont {Haake}}]{MKus_1988}%
  \BibitemOpen
  \bibfield  {author} {\bibinfo {author} {\bibfnamefont {M.}~\bibnamefont {Kus}}, \bibinfo {author} {\bibfnamefont {J.}~\bibnamefont {Mostowski}},\ and\ \bibinfo {author} {\bibfnamefont {F.}~\bibnamefont {Haake}},\ }\bibfield  {title} {\bibinfo {title} {Universality of eigenvector statistics of kicked tops of different symmetries},\ }\href {https://doi.org/10.1088/0305-4470/21/22/006} {\bibfield  {journal} {\bibinfo  {journal} {Journal of Physics A}\ }\textbf {\bibinfo {volume} {21}},\ \bibinfo {pages} {L1073} (\bibinfo {year} {1988})}\BibitemShut {NoStop}%
\bibitem [{\citenamefont {Izrailev}(1987)}]{izrailev1987chaotic}%
  \BibitemOpen
  \bibfield  {author} {\bibinfo {author} {\bibfnamefont {F.~M.}\ \bibnamefont {Izrailev}},\ }\bibfield  {title} {\bibinfo {title} {Chaotic stucture of eigenfunctions in systems with maximal quantum chaos},\ }\href@noop {} {\bibfield  {journal} {\bibinfo  {journal} {Physics Letters A}\ }\textbf {\bibinfo {volume} {125}},\ \bibinfo {pages} {250} (\bibinfo {year} {1987})}\BibitemShut {NoStop}%
\bibitem [{\citenamefont {Atas}\ \emph {et~al.}(2013)\citenamefont {Atas}, \citenamefont {Bogomolny}, \citenamefont {Giraud},\ and\ \citenamefont {Roux}}]{atas2013distribution}%
  \BibitemOpen
  \bibfield  {author} {\bibinfo {author} {\bibfnamefont {Y.~Y.}\ \bibnamefont {Atas}}, \bibinfo {author} {\bibfnamefont {E.}~\bibnamefont {Bogomolny}}, \bibinfo {author} {\bibfnamefont {O.}~\bibnamefont {Giraud}},\ and\ \bibinfo {author} {\bibfnamefont {G.}~\bibnamefont {Roux}},\ }\bibfield  {title} {\bibinfo {title} {Distribution of the ratio of consecutive level spacings in random matrix ensembles},\ }\href {https://doi.org/10.1103/PhysRevLett.110.084101} {\bibfield  {journal} {\bibinfo  {journal} {Phys. Rev. Lett.}\ }\textbf {\bibinfo {volume} {110}},\ \bibinfo {pages} {084101} (\bibinfo {year} {2013})}\BibitemShut {NoStop}%
\bibitem [{\citenamefont {Van~der Vaart}(2000)}]{van2000asymptotic}%
  \BibitemOpen
  \bibfield  {author} {\bibinfo {author} {\bibfnamefont {A.~W.}\ \bibnamefont {Van~der Vaart}},\ }\href@noop {} {\emph {\bibinfo {title} {Asymptotic Statistics}}},\ Vol.~\bibinfo {volume} {3}\ (\bibinfo  {publisher} {Cambridge University Press, Cambridge, UK},\ \bibinfo {year} {2000})\BibitemShut {NoStop}%
\bibitem [{\citenamefont {Luca~D'Alessio}\ and\ \citenamefont {Rigol}(2016)}]{doi:10.1080/00018732.2016.1198134}%
  \BibitemOpen
  \bibfield  {author} {\bibinfo {author} {\bibfnamefont {A.~P.}\ \bibnamefont {Luca~D'Alessio}, \bibfnamefont {Yariv~Kafri}}\ and\ \bibinfo {author} {\bibfnamefont {M.}~\bibnamefont {Rigol}},\ }\bibfield  {title} {\bibinfo {title} {From quantum chaos and eigenstate thermalization to statistical mechanics and thermodynamics},\ }\href {https://doi.org/10.1080/00018732.2016.1198134} {\bibfield  {journal} {\bibinfo  {journal} {Advances in Physics}\ }\textbf {\bibinfo {volume} {65}},\ \bibinfo {pages} {239} (\bibinfo {year} {2016})},\ \Eprint {https://arxiv.org/abs/https://doi.org/10.1080/00018732.2016.1198134} {https://doi.org/10.1080/00018732.2016.1198134} \BibitemShut {NoStop}%
\bibitem [{\citenamefont {Feingold}\ and\ \citenamefont {Peres}(1986)}]{PhysRevA.34.591}%
  \BibitemOpen
  \bibfield  {author} {\bibinfo {author} {\bibfnamefont {M.}~\bibnamefont {Feingold}}\ and\ \bibinfo {author} {\bibfnamefont {A.}~\bibnamefont {Peres}},\ }\bibfield  {title} {\bibinfo {title} {Distribution of matrix elements of chaotic systems},\ }\href {https://doi.org/10.1103/PhysRevA.34.591} {\bibfield  {journal} {\bibinfo  {journal} {Phys. Rev. A}\ }\textbf {\bibinfo {volume} {34}},\ \bibinfo {pages} {591} (\bibinfo {year} {1986})}\BibitemShut {NoStop}%
\bibitem [{\citenamefont {Wigner}(1955)}]{Wig55a}%
  \BibitemOpen
  \bibfield  {author} {\bibinfo {author} {\bibfnamefont {E.~P.}\ \bibnamefont {Wigner}},\ }\bibfield  {title} {\bibinfo {title} {Characteristic vectors of bordered matrices with infinite dimensions},\ }\href@noop {} {\bibfield  {journal} {\bibinfo  {journal} {Ann. Math.}\ }\textbf {\bibinfo {volume} {62}},\ \bibinfo {pages} {548} (\bibinfo {year} {1955})}\BibitemShut {NoStop}%
\bibitem [{\citenamefont {Casati}\ \emph {et~al.}(1980)\citenamefont {Casati}, \citenamefont {Guarneri},\ and\ \citenamefont {Valz-Gris}}]{conjectureCasati}%
  \BibitemOpen
  \bibfield  {author} {\bibinfo {author} {\bibfnamefont {G.}~\bibnamefont {Casati}}, \bibinfo {author} {\bibfnamefont {I.}~\bibnamefont {Guarneri}},\ and\ \bibinfo {author} {\bibfnamefont {F.}~\bibnamefont {Valz-Gris}},\ }\bibfield  {title} {\bibinfo {title} {On the connection between quantization of nonintegrable systems and statistical theory of spectra},\ }\href@noop {} {\bibfield  {journal} {\bibinfo  {journal} {Lett. Nuovo Cimento Soc. Ital. Fis.}\ }\textbf {\bibinfo {volume} {28}},\ \bibinfo {pages} {279} (\bibinfo {year} {1980})}\BibitemShut {NoStop}%
\bibitem [{\citenamefont {Guhr}\ \emph {et~al.}(1998)\citenamefont {Guhr}, \citenamefont {M\"{u}ller-Groeling},\ and\ \citenamefont {Weidenm\"{u}ller}}]{guhr98randomfull}%
  \BibitemOpen
  \bibfield  {author} {\bibinfo {author} {\bibfnamefont {T.}~\bibnamefont {Guhr}}, \bibinfo {author} {\bibfnamefont {A.}~\bibnamefont {M\"{u}ller-Groeling}},\ and\ \bibinfo {author} {\bibfnamefont {H.~A.}\ \bibnamefont {Weidenm\"{u}ller}},\ }\bibfield  {title} {\bibinfo {title} {Random matrix theories in quantum physics: Common concepts},\ }\href@noop {} {\bibfield  {journal} {\bibinfo  {journal} {Phys. Rep.}\ }\textbf {\bibinfo {volume} {299}},\ \bibinfo {pages} {189} (\bibinfo {year} {1998})},\ \Eprint {https://arxiv.org/abs/arXiv:cond-mat/9707301} {arXiv:cond-mat/9707301} \BibitemShut {NoStop}%
\bibitem [{\citenamefont {Pineda}\ and\ \citenamefont {Seligman}(2008)}]{pinedaseligmanELAF}%
  \BibitemOpen
  \bibfield  {author} {\bibinfo {author} {\bibfnamefont {C.}~\bibnamefont {Pineda}}\ and\ \bibinfo {author} {\bibfnamefont {T.~H.}\ \bibnamefont {Seligman}},\ }\bibfield  {title} {\bibinfo {title} {Random matrix models for decoherence and fidelity decay in quantum information systems},\ }in\ \href {https://doi.org/10.1063/1.2907760} {\emph {\bibinfo {booktitle} {LATIN-AMERICAN SCHOOL OF PHYSICS XXXVIII ELAF: Quantum Information and Quantum Cold Matter}}},\ Vol.\ \bibinfo {volume} {994}\ (\bibinfo  {publisher} {AIP},\ \bibinfo {year} {2008})\ pp.\ \bibinfo {pages} {47--75},\ \Eprint {https://arxiv.org/abs/arXiv:0711.1503} {arXiv:0711.1503} \BibitemShut {NoStop}%
\bibitem [{\citenamefont {Gorin}\ \emph {et~al.}(2008)\citenamefont {Gorin}, \citenamefont {Pineda}, \citenamefont {Kohler},\ and\ \citenamefont {Seligman}}]{Gorin_2008}%
  \BibitemOpen
  \bibfield  {author} {\bibinfo {author} {\bibfnamefont {T.}~\bibnamefont {Gorin}}, \bibinfo {author} {\bibfnamefont {C.}~\bibnamefont {Pineda}}, \bibinfo {author} {\bibfnamefont {H.}~\bibnamefont {Kohler}},\ and\ \bibinfo {author} {\bibfnamefont {T.~H.}\ \bibnamefont {Seligman}},\ }\bibfield  {title} {\bibinfo {title} {A random matrix theory of decoherence},\ }\href {https://doi.org/10.1088/1367-2630/10/11/115016} {\bibfield  {journal} {\bibinfo  {journal} {New Journal of Physics}\ }\textbf {\bibinfo {volume} {10}},\ \bibinfo {pages} {115016} (\bibinfo {year} {2008})}\BibitemShut {NoStop}%
\bibitem [{\citenamefont {Chavda}\ \emph {et~al.}(2014)\citenamefont {Chavda}, \citenamefont {Deota},\ and\ \citenamefont {Kota}}]{chavda2014poisson}%
  \BibitemOpen
  \bibfield  {author} {\bibinfo {author} {\bibfnamefont {N.}~\bibnamefont {Chavda}}, \bibinfo {author} {\bibfnamefont {H.}~\bibnamefont {Deota}},\ and\ \bibinfo {author} {\bibfnamefont {V.}~\bibnamefont {Kota}},\ }\bibfield  {title} {\bibinfo {title} {Poisson to {GOE} transition in the distribution of the ratio of consecutive level spacings},\ }\href@noop {} {\bibfield  {journal} {\bibinfo  {journal} {Physics Letters A}\ }\textbf {\bibinfo {volume} {378}},\ \bibinfo {pages} {3012} (\bibinfo {year} {2014})}\BibitemShut {NoStop}%
\bibitem [{\citenamefont {Casati}\ \emph {et~al.}(1996)\citenamefont {Casati}, \citenamefont {Chirikov}, \citenamefont {Guarneri},\ and\ \citenamefont {Izrailev}}]{CASATI1996430}%
  \BibitemOpen
  \bibfield  {author} {\bibinfo {author} {\bibfnamefont {G.}~\bibnamefont {Casati}}, \bibinfo {author} {\bibfnamefont {B.}~\bibnamefont {Chirikov}}, \bibinfo {author} {\bibfnamefont {I.}~\bibnamefont {Guarneri}},\ and\ \bibinfo {author} {\bibfnamefont {F.}~\bibnamefont {Izrailev}},\ }\bibfield  {title} {\bibinfo {title} {Quantum ergodicity and localization in conservative systems: the wigner band random matrix model},\ }\href {https://doi.org/https://doi.org/10.1016/S0375-9601(96)00784-0} {\bibfield  {journal} {\bibinfo  {journal} {Physics Letters A}\ }\textbf {\bibinfo {volume} {223}},\ \bibinfo {pages} {430} (\bibinfo {year} {1996})}\BibitemShut {NoStop}%
\bibitem [{\citenamefont {Kota}(2001)}]{kota2001embedded}%
  \BibitemOpen
  \bibfield  {author} {\bibinfo {author} {\bibfnamefont {V.}~\bibnamefont {Kota}},\ }\bibfield  {title} {\bibinfo {title} {Embedded random matrix ensembles for complexity and chaos in finite interacting particle systems},\ }\href@noop {} {\bibfield  {journal} {\bibinfo  {journal} {Physics Reports}\ }\textbf {\bibinfo {volume} {347}},\ \bibinfo {pages} {223} (\bibinfo {year} {2001})}\BibitemShut {NoStop}%
\bibitem [{\citenamefont {Trotter}(1959)}]{trotter1959product}%
  \BibitemOpen
  \bibfield  {author} {\bibinfo {author} {\bibfnamefont {H.~F.}\ \bibnamefont {Trotter}},\ }\bibfield  {title} {\bibinfo {title} {On the product of semi-groups of operators},\ }\href@noop {} {\bibfield  {journal} {\bibinfo  {journal} {Proceedings of the American Mathematical Society}\ }\textbf {\bibinfo {volume} {10}},\ \bibinfo {pages} {545} (\bibinfo {year} {1959})}\BibitemShut {NoStop}%
\bibitem [{\citenamefont {Suzuki}(1976)}]{suzuki1976generalized}%
  \BibitemOpen
  \bibfield  {author} {\bibinfo {author} {\bibfnamefont {M.}~\bibnamefont {Suzuki}},\ }\bibfield  {title} {\bibinfo {title} {Generalized trotter's formula and systematic approximants of exponential operators and inner derivations with applications to many-body problems},\ }\href@noop {} {\bibfield  {journal} {\bibinfo  {journal} {Communications in Mathematical Physics}\ }\textbf {\bibinfo {volume} {51}},\ \bibinfo {pages} {183} (\bibinfo {year} {1976})}\BibitemShut {NoStop}%
\bibitem [{\citenamefont {Suzuki}(1985)}]{suzuki1985decomposition}%
  \BibitemOpen
  \bibfield  {author} {\bibinfo {author} {\bibfnamefont {M.}~\bibnamefont {Suzuki}},\ }\bibfield  {title} {\bibinfo {title} {Decomposition formulas of exponential operators and lie exponentials with some applications to quantum mechanics and statistical physics},\ }\href@noop {} {\bibfield  {journal} {\bibinfo  {journal} {Journal of mathematical physics}\ }\textbf {\bibinfo {volume} {26}},\ \bibinfo {pages} {601} (\bibinfo {year} {1985})}\BibitemShut {NoStop}%
\bibitem [{\citenamefont {Rabinovici}\ \emph {et~al.}(2022{\natexlab{b}})\citenamefont {Rabinovici}, \citenamefont {S{\'{a}}nchez-Garrido}, \citenamefont {Shir},\ and\ \citenamefont {Sonner}}]{Rabinovici_2022_localization}%
  \BibitemOpen
  \bibfield  {author} {\bibinfo {author} {\bibfnamefont {E.}~\bibnamefont {Rabinovici}}, \bibinfo {author} {\bibfnamefont {A.}~\bibnamefont {S{\'{a}}nchez-Garrido}}, \bibinfo {author} {\bibfnamefont {R.}~\bibnamefont {Shir}},\ and\ \bibinfo {author} {\bibfnamefont {J.}~\bibnamefont {Sonner}},\ }\bibfield  {title} {\bibinfo {title} {Krylov localization and suppression of complexity},\ }\bibfield  {journal} {\bibinfo  {journal} {Journal of High Energy Physics}\ }\textbf {\bibinfo {volume} {2022}},\ \href {https://doi.org/10.1007/jhep03(2022)211} {10.1007/jhep03(2022)211} (\bibinfo {year} {2022}{\natexlab{b}})\BibitemShut {NoStop}%
\bibitem [{\citenamefont {Rabinovici}\ \emph {et~al.}(2023)\citenamefont {Rabinovici}, \citenamefont {S{\'a}nchez-Garrido}, \citenamefont {Shir},\ and\ \citenamefont {Sonner}}]{rabinovici2023bulk}%
  \BibitemOpen
  \bibfield  {author} {\bibinfo {author} {\bibfnamefont {E.}~\bibnamefont {Rabinovici}}, \bibinfo {author} {\bibfnamefont {A.}~\bibnamefont {S{\'a}nchez-Garrido}}, \bibinfo {author} {\bibfnamefont {R.}~\bibnamefont {Shir}},\ and\ \bibinfo {author} {\bibfnamefont {J.}~\bibnamefont {Sonner}},\ }\bibfield  {title} {\bibinfo {title} {A bulk manifestation of krylov complexity},\ }\href@noop {} {\bibfield  {journal} {\bibinfo  {journal} {Journal of High Energy Physics}\ }\textbf {\bibinfo {volume} {2023}},\ \bibinfo {pages} {1} (\bibinfo {year} {2023})}\BibitemShut {NoStop}%
\bibitem [{\citenamefont {Mirkin}\ and\ \citenamefont {Wisniacki}(2021)}]{PhysRevE.103.L020201}%
  \BibitemOpen
  \bibfield  {author} {\bibinfo {author} {\bibfnamefont {N.}~\bibnamefont {Mirkin}}\ and\ \bibinfo {author} {\bibfnamefont {D.}~\bibnamefont {Wisniacki}},\ }\bibfield  {title} {\bibinfo {title} {Quantum chaos, equilibration, and control in extremely short spin chains},\ }\href {https://doi.org/10.1103/PhysRevE.103.L020201} {\bibfield  {journal} {\bibinfo  {journal} {Phys. Rev. E}\ }\textbf {\bibinfo {volume} {103}},\ \bibinfo {pages} {L020201} (\bibinfo {year} {2021})}\BibitemShut {NoStop}%
\end{thebibliography}%
\end{document}